\documentclass[allclo]{FBSart}
\usepackage{epsfig}
\newcommand{\ga}{\alpha}
\newcommand{\gb}{\beta}
\newcommand{\gc}{\gamma}
\newcommand{\gd}{\delta}

\newcommand{\gl}{\lambda}

\newcommand{\hh}{\frac{1}{2}}

\title{ Light-front field theory of hot and dense quark
  matter\footnote{Presented at Light-Cone 2004, Amsterdam, 16 - 20
    August}}
\author{Michael Beyer}
\institute{Fachbereich Physik, Universit\"at Rostock, 18051 Rostock, Germany\\
  email: {\tt michael.beyer@uni-rostock.de}}
\runningauthor{Michael Beyer}
\runningtitle{LC 2004}
\begin{document}
\maketitle
\begin{abstract}
  Extending the concepts of light-front field theory to quantum statistics
  provides a novel approach towards nuclear matter under extreme conditions.
  Such conditions exist, e.g., in neutron stars or in the early stage of our
  universe.  They are experimentally expected to occur in heavy ion
  collisions, e.g., at RHIC and accelerators to be build at GSI and CERN.
  Light-front field theory is particularly suited, since it is based on a
  relativistic Hamiltonian approach. It allows us to treat the perturbative as
  well as the nonperturbative regime of QCD and also correlations that emerge
  as a field of few-body physics and is important for hadronization.  Last but
  not least the Hamiltonian approach is useful for nonequilibrium processes by
  utilizing, e.g., the formalism of nonequilibrium statistical operators.
\end{abstract}
\section{Finite temperature and the light-front}
At first sight, it seems difficult to define proper temperature $T$ for a
system ``on the light-front'', because there is no Lorentz transformation to
the rest frame of an observer, who is holding the thermometer. Despite the
fact that this is formally clarified by now (see Sect. 2), I would like to
motivate, how temperature could survive ``on the light-front''. Consider the
Fermi distribution for the canonical ensemble of a noninteracting gas in the
instant form
\begin{equation}
f(k)=\big[\exp(k^0_{\rm on}/T) +1\big]^{-1},\label{eqn:fermi}
\end{equation}
where $k^0_{\rm on}=+\sqrt{m^2+\vec k^2}$ is the on-shell energy of a
one-particle state with respect to the medium frame.  Let the velocity of the
medium be given by the time-like vector $u=(u^0,\vec u)$, where $u^2=1$. If
the medium is at rest with respect to the observer, $u=(1,\vec 0)$. For
convenience we introduce the momentum of the medium\footnote{This can be done,
  e.g., before the thermodynamic limit, so that the mass of the noninteracting
  system $M_0$ can be achieved by summing all the single-particle energies.}
$P=M_0 u$ with $P^2=M_0^2$.  Following Susskind~\cite{Sus} we now assume an
observer moving with a large velocity $v$ along the $z$ axis in the negative
$z$ direction. The one-particle momentum components seen from this moving
frame are denoted $p^\mu$. The momentum of the medium measured with respect to
this system is $P^\mu=(E,0,0,P)$ with $E=\sqrt{M_0^2+P^2}$. As long as $P<\infty$
the relation between the medium system and the observer's system are given by
Lorentz transformations $L(\omega_P)$. For
$P\rightarrow\infty$ we introduce the momentum fraction $\eta$ via $p^3=\eta
P$ and expand $E$ and $p^0_{\rm on}$, hence $L(\omega_P)$, up to ${\cal
  O}(P^{-2})$ and finally arrive at (in the limit $P\rightarrow \infty$)
\begin{equation}
k^0=\frac{1}{2}\left(\eta M_0+\frac{{\vec p}_\perp^2+m^2}{\eta M_0}\right),\quad
k^3=\frac{1}{2}\left(\eta M_0-\frac{{\vec p}_\perp^2+m^2}{\eta M_0}\right),\quad
\vec k_\perp = \vec p_\perp.\label{eqn:imf}
\end{equation}
This form suggests to introduce light-cone components and to identify
the longitudinal fraction $\eta$ with the (kinematically invariant)
light-cone momentum fraction $x=p^+/P^+=k^+/M_0$. After doing so
(\ref{eqn:imf}) becomes the well known transformation of momentum
variables~\cite{Sus},
\begin{equation}
k^0=\frac{1}{2}\left(k^+-k^-\right),\quad
k^3=\frac{1}{2}\left(k^+-k^-\right),\quad
\vec k_\perp = \vec p_\perp.\label{eqn:lf}
\end{equation}
Hence the Fermi distribution ``on the light-front'' is given by
\begin{equation}
f(k)=\left[\exp\left(\frac{k^-_{\rm on}+k^+}{2T}\right) +1\right]^{-1},
\label{eqn:fermiLC}
\end{equation}
where the on-shell condition in (\ref{eqn:fermi}) is evaluated for the
light-front coordinates, viz.  $k^-_{\rm on}=(m^2+\vec k^2_\perp)/k^+$.
Despite the fact that we have introduced light-front coordinates and done a
limiting process $P\rightarrow\infty$ the components of the four vector
$k^\mu$ that appear in (\ref{eqn:fermiLC}) are given with respect to the {\em
  medium frame}. Also the value for $T$ is still given with respect to the
{\em medium frame}. This is how temperature is usually defined. The
thermometer must be in contact with the medium ``for a long time'' (sometimes
called zero's law of thermodynamics).  No Lorentz transformation is now
possible to the rest system of an observer.  It is not even necessary to
connect the temperature to a rest frame of an ``{\em instant form observer}''.
The medium frame is relevant. This reflects precisely the different forms of
dynamics first introduced by Dirac~\cite{Dir}.  Of course arbitrary Lorentz
transformations of the medium frame to different frames are still possible and
hence a covariant formalism can be achieved also in the light-front form.

\section{Statistical physics in the light-front quantization}

A formal framework of covariant calculations at finite temperatures {\em in
  instant form} has been given in Ref.~\cite{Israel:1976tn}
in a different context. The grand
canonical partition operator is given by
\begin{equation}
Z_G =  \exp\left\{-(u\cdot P  - \mu N)/T\right\},
\label{eqn:ZG}
\end{equation}
where $\mu$ the chemical potential. The statistical operator is
$\rho_G=Z_G/{\rm Tr} Z_G$. It is clear that (\ref{eqn:fermi}) can be achieved
by choosing $u^\nu=(1,\vec 0)$, which are the components of $u$ as seen from
the medium frame (and, in instant form, the observer's rest frame).  In
light-front quantization we give up the connection between observer's rest
frame and medium frame. As we have seen in the introduction, giving the
components of the velocity in the medium frame suffice. We shall arrive at
(\ref{eqn:fermiLC}) choosing
\begin{equation}
u^\nu = (u^+,u^-,\vec u^{\perp})= (1,1,0,0).
\label{eqn:vel}
\end{equation} 
Following notation of
Ref.~\cite{Brodsky:1997de} the operators appearing in (\ref{eqn:ZG}) are given by
\begin{equation}
P^\mu=\int d\omega_+\; T^{+\mu}(x),\quad N=\int d\omega_+\; J^+(x),
\label{eqn:P}
\end{equation}
where $T^{\nu\mu}(x)$ denotes the energy momentum tensor, $J^\nu(x)$ is the
conserved current, and $d\omega_+=dx_-d^2x_\perp$. With the above choice of $u$ the
partition operator can then be written as
\begin{equation}
Z_G  =\exp\left\{-\frac{1}{T}\left(\frac{1}{2}P^-+\frac{1}{2}P^+ 
- \mu N\right)\right\}.
\end{equation}
The resulting Fermi distribution functions of particles $f^+$ and
antiparticles $f^-$ are then given
by~\cite{Beyer:2001bc,Mattiello:2001vq}
\begin{eqnarray}
f^\pm(k)&=&
\left[\exp\left\{\left(u\cdot k_\mathrm{on}
\mp\mu\right)/T\right\}+1\right]^{-1}
\label{eqn:fermipm}
\end{eqnarray}
with $k_{\mathrm{on}}=(k^-_{\mathrm{on}},k^+,\vec k_\perp)$ and $u\cdot
k_\mathrm{on}=(k^-_{\mathrm{on}}+k^+)/2$.

\section{Many-body Green functions}
To treat hot and dense quantum systems, such as nuclear or quark matter in
question here, we use the techniques of many-body Green functions.  Reference
to textbook treatments is given in~\cite{fet71}. Many-body Green functions,
even in the nonrelativistic case, are complicated objects.  We organize the
equations in a Dyson expansion that leads to a hierarchy of linked cluster
equations. The Dyson equation approach to correlations in many-body systems
along with several approximation schemes has been developed by Schuck and
collaborators. A review and references are given in~\cite{duk98}.  In recent
years we have systematically applied this approach for finite temperatures up
to multi-nucleon clusters (see e.g.~\cite{Beyer:1996rx} and refs therein).
Here I give a brief summary and the generalization to the light-front
dynamics~\cite{Beyer:2001bc,Mattiello:2001vq}.  In
fact, the light-cone quantization is well suited to use the many-body Green
function formalism as it utilizes the Fock space
expansion~\cite{Brodsky:1997de}. We define a light-cone time-ordered Green
function for a given number of particles as follows (retarded and advanced
many-body Green functions can be defined accordingly)
\begin{eqnarray} 
      {\cal G}_{\alpha\beta}^{x^+-x'^+}&=&-i\langle T_+ A_\ga(x^+)
          A_\gb^\dagger(x'^+)\rangle\label{eqn:green}\\
       &=&-i\left(\theta(x^+-x'^+)\langle A_\gb^\dagger(x'^+)A_\ga(x^+)
      \rangle\mp\theta(x'^+-x^+)\langle A_\gb^\dagger(x'^+)A_\ga(x^+)
       \rangle\right).\nonumber
\end{eqnarray}
The average $\langle\cdots\rangle$ is taken over the exact ground state and
the upper (lower) sign is for fermions (bosons). The operators $A_\ga(x^+)$
could be build out of any number of field operators (fermions and/or bosons).
The light-cone time dependence of the operators is given in the Heisenberg picture by
$A(x^+)=e^{iP_+x^+}Ae^{-iP_+x^+}$. 
For $A(x^+)=\Psi(x^+)$, the free Fermion
field, the standard light-front propagator is recovered.

In finite temperature formalism the above definition can be generalized.  For
a grand canonical ensemble on the light-front, the generalized Heisenberg
picture assumes the form $$A(\tau)=e^{H\tau}Ae^{-H\tau},$$
where $H=u\cdot
P-\mu N$. Note however that $H$ is not the bare Hamiltonian $P_+$, but
includes additional terms related to Lagrange constraints.  Following the
convention of~\cite{fet71} the respective Green function is defined by
\begin{equation} 
    {\cal G}_{\alpha\beta}^{\tau-\tau'}
        =-\langle T_\tau A_\ga(\tau)A_\gb^\dagger(\tau')\rangle.
\end{equation}
The average is now taken over the (equilibrium) grand canonical statistical
operator $\rho_G$, viz.  $\langle\cdots\rangle={\rm tr}\{\rho_G\dots\}$. For
$\tau= ix^+$ (``imaginary time'') the Heisenberg picture is formally
recovered. The ``true''~\cite{fet71} Heisenberg picture with respect
to this Hamiltonian $H$ is given by $$A(x^+)=e^{iHx^+}Ae^{-iHx^+}.$$
Using
this operator in (\ref{eqn:green}) defines the ``real time'' many-body Green
function.
Dyson equations can be established for both forms~\cite{duk98,bey97}. In the
light-front formalism the Dyson equations are given by~\cite{Beyer:2001bc}
\begin{equation} 
    i\frac{\partial}{\partial{x^+}}\; {\cal G}^{{x^+}-{x'^+}}_{\ga\gb}
     =\delta({x^+}-{x'^+}) \langle [A_\ga,A_\gb^\dagger]_\pm\rangle
    + \sum_{\gc}\int d{\bar x^+}\; {{\cal M}^{{x^+}- {\bar x^+}}_{\ga\gc}}
    \;{\cal G}^{{\bar x^+}-{x'^+}}_{\gc\gb}.
\label{eqn:dyson}
\end{equation}
The mass matrix that appears in (\ref{eqn:dyson}) is given by
\begin{eqnarray} 
        {\cal M}^{{x^+}-{x'^+}}_{\ga\gb}
        &=& \delta({x^+}-{x'^+}) {\cal M}^{{x^+}}_{0,\ga\gb}
        + {\cal M}^{{x^+}-{x'^+}}_{r,\ga\gb}\label{eqn:mass}\\
       ( {\cal M}_0{\cal N})^{x^+}_{\ga\gb}&=
        & \langle [[A_\ga,H]({x^+}),A_\gb^\dagger({x^+})]_\pm\rangle
\label{eqn:mass0}\\
        ({\cal M}_r{\cal N})^{{x^+}-{x'^+}}_{\alpha\beta}&=&
       \sum_\gamma\langle T_{x^+} [A_\alpha,H]({x^+}),
       [A^\dagger_\gb,H]({x'^+})]\rangle_{{\rm irreducible}}
\end{eqnarray} 
where ${\cal
  N}^{x^+}_{\ga\gb}=\langle[A_\ga,A_\gb^\dagger]_\pm({x^+})\rangle$.  The
first term in (\ref{eqn:mass}) is instantaneous and related to the mean field
approximation, the second term is the retardation or memory term.  This form
suggests to first solve the mean field problem (neglecting memory) and
evaluate higher order contributions involving ${\cal
  M}^{{x^+}-{x'^+}}_{r,\ga\gb}$ in the mean field basis.  In this
approximation (\ref{eqn:dyson}) can be written in the following form
\begin{equation} 
      \left(  i\frac{\partial}{\partial{x^+}}\;
-  {{\cal M}^{{x^+}}_{0}}\right){\cal G}^{{x^+}-{x'^+}}
        =\delta({x^+}-{x'^+}) {\cal N}^{x^+},\label{eqn:mf}
\end{equation}
where ${\cal M}$, ${\cal G}$ etc. are understood as matrices acting in the
index space of $\ga,\gb\,\dots$ As an example I give results for the ideal gas (of
particles or quasiparticles).  Following (almost) the notation
of~\cite{Brodsky:1997de} the Fock space representation of $H$ is given by
\begin{equation} 
H=\sum\int dp^+d^2p_\perp \left[(u\cdot p -\mu) b^\dagger b +
 (u\cdot p + \mu) d^\dagger d\right],
\end{equation}
where the sum is over all quantum numbers relevant for the particles $b$ or
antiparticles $d$. It is now straight forward to evaluate ${\cal N}$ and
${\cal M}_0$ using free fermion field operators in (\ref{eqn:green}). The
resulting single-particle light-cone time ordered Green function (ideal gas)
is 
\begin{eqnarray}
G(k)&=&
\frac{\gamma k_\mathrm{on}+m}{\hh z-\hh k^-_\mathrm{on}+i\varepsilon}
 \frac{\theta (k^+)}{2k^+}
(1-f^+)
+\frac{\gamma k_\mathrm{on}+m}{\hh z-\hh  k^-_\mathrm{on}-i\varepsilon} 
\frac{\theta(k^+)}{2k^+}
f^+\label{eqn:lfmed}\\
&&
+\frac{\gamma k_\mathrm{on}+m}{\hh z-\hh  k^-_\mathrm{on}+i\varepsilon} 
\frac{\theta(-k^+)}{2k^+}
f^-
+\frac{\gamma k_\mathrm{on}+m}{\hh z-\hh k^-_\mathrm{on}-i\varepsilon} 
\frac{\theta(-k^+)}{2k^+}
(1-f^-).
\nonumber
\end{eqnarray}
were we used (\ref{eqn:vel}) and $\hh z=\hh k^- +\mu -\hh k^+$ for
convenience. The ensemble averages needed are
\begin{equation}
\langle b^\dagger b\rangle = f^+,\quad
\langle b b^\dagger\rangle =1- f^+,\quad
\langle d^\dagger d\rangle = f^-, \quad
\langle d d^\dagger\rangle =1- f^-.
\end{equation}
Eq. (\ref{eqn:lfmed}) is the light-front generalization of equation (31.38)
of~\cite{fet71}, which is connected to the Matsubara-Fourier representation via
the generalized Lehmann representation~\cite{fet71}.  The Matsubara
frequencies analytically continued to $\hh z$ are given by~\cite{Beyer:2001bc}
\begin{equation}
     \hh k^-_n = \left\{\begin{array}{ll} i(2n+1)\pi T +\mu -\hh k^+&{\rm fermions,}\\[1ex]
                            i2n\pi T +\mu -\hh k^+&{\rm bosons.}\end{array}\right.
\end{equation}
In Matsubara representation the $i\varepsilon$ term can be dropped in
(\ref{eqn:lfmed}), hence it is obvious that the Pauli factors $f^\pm$ vanish
(which is only valid for the one-particle Green function) and the ``imaginary
time'' Green function sums up to
\begin{equation}
G(k_n)=
\frac{\gamma k_{\mathrm{on}}+m}{k_n^2-m^2}
\label{eqn:Gim}
\end{equation}
where $k_n=(k^-_n,k^+,\vec k_\perp)$, which have been given in similar form
in~\cite{Alves:2002tx}. Our light-front in-medium Green function
(\ref{eqn:lfmed}) given first in \cite{Beyer:2001bc} shows no basic difference
(up to conventions and phase space) from the ones given
later~\cite{Alves:2002tx}.  However, we do have used a different representation
more useful for actual few-body calculations.
Similar evaluation of two- or three-body Fock states leads to the respective
two-particle or three-particle equations for the two- and three-body
system embedded in a hot and dense medium. The proper Pauli blocking factors
follow straight forwardly.

\section{Conclusion and prospects}

Generalizing many-body Green functions~\cite{fet71} to the light-front and
utilizing a cluster expansion we are able to treat few-body correlations in a
hot and dense (i.e. relativistic) plasma. As a model we have used an effective
zero range interaction~\cite{mattiello}. This approach provides the framework
to tackle such intriguing questions how hadrons form during the early plasma
phase of the universe, if there is a color superconducting phase, what is the
nature of QCD phase transition. Also, because of the Hamiltonian form of the
light-front quantization, it is possible to invoke the whole formalism of
nonequilibrium statistical physics and derive Kadanov-Baym or Boltzmann
equations that could be used to describe the relativistic transport during a
heavy ion collision. It is also possible (and necessary) to go beyond the zero
range model and implement real light-front QCD, which has been elaborated in
the works by Brodsky, Pauli and Pinsky~\cite{Brodsky:1997de}.

{\em Acknowledgment:} I would like to thank the organizers for the pleasant
and inspiring atmosphere during this meeting. I am grateful to H.J. Weber and
T. Frederico for all the fruitful discussions and collaboration on the topic.
Also I would like to thank S. Mattiello and S. Strau{\ss} for their contributions
to this work. This work is supported by Deutsche Forschungsgemeinschaft.
\end{document}